\newcommand{\EQ}{\begin{equation}}
\newcommand{\EE}{\end{equation}}
\newcommand{\EQA}{\begin{eqnarray}}
\newcommand{\EEA}{\end{eqnarray}}
\newcommand{\brac}[1]{\langle #1 \rangle}

\newcommand{\etatz}{\eta_{\rm t0}}

\newcommand{\urms}{u_{\rm rms}}

\newcommand{\Beq}{B_{\rm eq}}

\newcommand{\Rm}{R_{\rm m}}

\newcommand{\BB}{{\bm{B}}}
\newcommand{\bb}{{\bm{b}}}
\newcommand{\JJ}{{\bm{J}}}
\newcommand{\jj}{{\bm{j}}}

\newcommand{\nab}{\mbox{\boldmath $\nabla$} {}}
{}

\def \grad {{\bm \nabla}}
\def \curl {{\bm \nabla} \times}
\def \dive {{\bm \nabla}\cdot}
\def \lap {\nabla^2}
\def \delt {\partial_t}
\def \Dt {D_t}

\def \u {{\bm U}}

\def \uu {{\bm u}}

\def \AA {{\bm A}}
\def \aa {{\bm a}}
\def \BB {{\bm B}}

\def \EE {{\bm E}}
\def \ee {{\bm e}}
\def \fxt {{\bm f}({\bm x},t)}
\def \f {{\bm f}}
\def \kf  {k_{\rm f}}

\def \urms  {u_{\rm rms}}

\def \Beq {B_{\rm eq}}
 
\def \hM {h^{\rm M}}
\def \HM {H^{\rm M}}

\def \hMM {\overline{h}^{\rm M}}
\def \hMm {\overline{h}^{\rm M}_{\rm m}}
\def \hMf {\overline{h}^{\rm M}_{\rm f}}

\def \Fh {{\cal F}^{\rm H}}
\def \Fhm {{\overline{\cal F}}^{\rm H}_{\rm m}}
\def \Fhf {{\overline{\cal F}}^{\rm H}_{\rm f}}
\def \FFh {{\mbox{\boldmath ${\cal F}$}}^{\rm H}}
\def \FFhh {\overline{\mbox{\boldmath ${\cal F}$}}^{\rm H}}
\def \FFhm {\overline{\mbox{\boldmath ${\cal F}$}}^{\rm H}_{\rm m}}
\def \FFhf {\overline{\mbox{\boldmath ${\cal F}$}}^{\rm H}_{\rm f}}
\def \meanEMF {\overline{\mbox{\boldmath ${\cal E}$}}}

\def \kappaf {\kappa_{\rm f}}
\newcommand{\dd}{{\rm d} {}}
\newcommand{\bra}[1]{\langle #1\rangle}
\newcommand{\oo}{\vec{\omega}}
\newcommand{\SSSS}{\mbox{\boldmath ${\sf S}$} {}}

\documentclass[mathleft]{an}
\usepackage{graphicx}
\usepackage{times}
\usepackage{bm}
\begin{document}
\graphicspath{{./fig/}{./png/}}
\sloppy

\title{Equatorial magnetic helicity flux in simulations with different gauges}

\author{Dhrubaditya Mitra\inst{1}\fnmsep\thanks{Corresponding author:
    \email{dhruba.mitra@gmail.com}}
     \and Simon Candelaresi\inst{2,3} \and Piyali Chatterjee\inst{2} 
     \and Reza Tavakol\inst{1}
     \and A. Brandenburg\inst{2,3}
}

\titlerunning{Equatorial magnetic helicity flux}
\authorrunning{D. Mitra et al.}

\institute{Astronomy Unit, School of Mathematical Sciences, Queen Mary
University of London, Mile End Road, London E1 4NS, United Kingdom
\and
NORDITA, AlbaNova University Center, Roslagstullsbacken 23, SE-10691 
Stockholm, Sweden
\and
Department of Astronomy, AlbaNova University Center,
Stockholm University, SE-10691 Stockholm, Sweden}


\date{Received 2009 Nov 4; $ $Revision: 1.129 $ $}


\abstract{
We use direct numerical 
simulations of forced MHD turbulence with a forcing function that produces
two different signs of kinetic helicity in the upper and lower parts of the domain.
We show that the mean flux of magnetic helicity from the small-scale field
between the two parts of the domain
can be described by a Fickian diffusion law with a diffusion coefficient
that is approximately independent of the magnetic Reynolds number and
about one third of the estimated turbulent magnetic diffusivity. 
The data suggest that the turbulent diffusive magnetic helicity flux can
only be expected to alleviate catastrophic quenching at Reynolds numbers
of more than several thousands.
We further calculate the magnetic helicity density and its flux
in the domain for three different gauges.
We consider the Weyl gauge, in which the electrostatic potential vanishes,
the pseudo-Lorenz gauge, where the speed of light is replaced by the sound
speed, and the `resistive gauge' in which the
Laplacian of the magnetic vector potential acts as resistive term.
We find that, in the statistically steady state, the time-averaged
magnetic helicity density and the magnetic helicity flux are the same
in all three gauges.
}

\maketitle

\section{Introduction}
\label{sec:intro}

The generation of magnetic fields on scales larger than the 
eddy scale of the underlying turbulence in astrophysical bodies has posed
a major problem.  
Magnetic helicity is believed to play an important
role in this process \cite{BS05}. The magnetic helicity density, defined by 
$\AA\cdot\BB$, where $\BB = \curl \AA$ is the magnetic field and
$\AA$ is the corresponding magnetic vector potential,
is important because at large scales
it is produced in many dynamos.
This has been demonstrated for dynamos based on the $\alpha$ effect
(\cite{SSSB06,BCC09}),
the shear--current effect \cite{BS05c}, and the incoherent $\alpha$--shear
effect \cite{BRRK08}.
The volume integral of the magnetic helicity density over periodic domains 
(as well as domains with perfect-conductor boundary
conditions or infinite domains where the magnetic field and the vector
potential decays fast enough at infinity) is a conserved quantity in 
ideal MHD. 
This conservation is also believed to be recovered in the limit
of infinite magnetic Reynolds number in non-ideal MHD \cite{Berger84}.
This implies that for finite (but large) magnetic Reynolds numbers
magnetic helicity can decay only through 
microscopic resistivity.
This would in turn control the saturation time and cycle periods
of large-scale {\it helical} magnetic field which would be 
too slow to explain the observed variations of magnetic
fields in astrophysical settings, such as for example
the 11 year variation of the large-scale fields during the solar cycle.

A possible way out of this deadlock is provided by
fluxes of magnetic helicity out of the domain (\cite{BF00,KMRS00}).
In the case of solar dynamo, such a flux could be out of the domain,
mediated by coronal mass ejections, or it could be across the equator,
mediated by internal fluxes within the domain.
Several possible candidates for magnetic helicity fluxes have been proposed
(\cite{KR99,vis+cho01,SB04}).

In this paper we measure the diffusive flux across the domain 
with two different signs of magnetic helicity. 
This measurement however poses an additional difficulty,
due to the fact that neither the flux nor the magnetic helicity 
density remain invariant under the gauge transformation
$\AA\rightarrow\AA+\nab\Lambda$, up to which the vector potential is defined.
This constitutes a gauge problem. 
This problem, however, does not arise in homogeneous (or nearly homogeneous)
domains with periodic or
perfect-conductor boundary conditions, or in infinitely large domains
where both the magnetic field and vector potential decay fast 
enough at infinity.
In these cases the volume integral
of magnetic helicity is {\it gauge-invariant}, because surface terms
vanish and $\nab\cdot\BB=0$, so that
$\int\BB\cdot\nab\Lambda\,\dd V=-\int\Lambda\nab\cdot\BB\,\dd V=0$.
However, in practice we are often interested in finite or open domains with 
more realistic boundary conditions. Also, if we are to talk meaningfully
about the exchange of magnetic helicity between two parts of the domain
we need to evaluate changes in magnetic helicity densities locally
even if the integral of the magnetic helicity density over the whole
domain is gauge-invariant.
An important question then is how to calculate this quantity
across arbitrary surfaces in numerical simulations. 
Ideally one would like to have a gauge-invariant description of magnetic 
helicity. 
A number of suggestions have been put forward 
in the literature (\cite{ber+fie84,SB06}).
In practice, however, calculating the gauge-invariant volume integral
of magnetic helicity poses an awkward complication and may not
be the quantity relevant for dynamo quenching (\cite{SB06}).
In this paper, to partially address this question, we take an alternative 
view and try to compare and contrast the magnetic helicity and its flux 
across the domain
in three different gauges that are often used in numerical simulations. 

\section{Model and Background}
\label{model}
The setup in this paper is  inspired by the recent work of
\cite{mit+tav+kap+bra09}, who
considered a wedge-shaped domain encompassing parts of both the
southern and northern hemispheres. Direct numerical simulations (DNS)
of the compressible MHD equations with an external force which 
injected negative (positive) helicity in the northern (southern) hemispheres
shows a dynamo with polarity reversals, oscillations and equatorward
migration of magnetic activity.  It was further shown, using mean-field
models, that such a dynamo is well described by an $\alpha^2$ dynamo,
where $\alpha$ has positive (negative) sign in the northern (southern) hemisphere. 
However, the mean-field dynamo showed catastrophic quenching,
i.e., the ratio of magnetic energy to the equipartition magnetic energy 
decreases as $\Rm^{-1}$, where $\Rm$ is the magnetic Reynolds number.
Such catastrophic quenching could
potentially be alleviated by a mean flux of small-scale magnetic helicity
across the equator (\cite{BCC09}).
Diffusive flux of this kind has previously been employed in 
mean-field models on empirical grounds (\cite{CTTB98,KMRS00}).
Using a one-dimensional mean-field  model of an $\alpha^2$ dynamo 
with positive $\alpha$ in the north and negative in the south,
it was possible to show that for large enough values of $\Rm$
catastrophic quenching is indeed alleviated (\cite{BCC09}).
However, three questions still remained:
\begin{enumerate}
\item  Can such a diffusive flux result from DNS? 
\item Is it strong enough to alleviate catastrophic quenching?
\item When is it independent of the gauge chosen?
\end{enumerate} 
In this paper we provide partial answers to these questions. 

We proceed by simplifying our problem further, both 
conceptually and numerically,
by considering simulations performed in a rectangular Cartesian box with dimensions
$L_x\times L_y\times L_z$.  The box is divided into two
equal cubes along the $z$ direction, with sides $L_x=L_y=L_z/2$. We shall
refer to the $xy$ plane at $z=0$  as the `equator' and the 
regions with positive (negative) $z$ as `north' and `south' respectively. 
We shall choose the helicity of the external 
force such that it has negative (positive) helicity in the 
northern (southern) parts of the domain. 
All the sides of the simulation 
domain are chosen to have periodic boundary conditions.
The slowest resistive decay rate of the mean magnetic field is
$\eta k_1^2$, where $\eta$ is the microscopic magnetic diffusivity
and $k_1=\pi/L_z$ is the lowest wavenumber of the domain.

We employ two different random forcing functions: one where the helicity
of the forcing function varies sinusoidally with $z$ (Model~A) and one
where it varies linearly with $z$ (Model~B).
This also leads to a corresponding variation of the kinetic and 
small-scale current helicities in the domain.
Model~A minimizes the possibility of boundary effects, while Model~B employs
the same profile as that used in an earlier mean-field model (\cite{BCC09}).
The typical wavenumber of the forcing function is chosen to be $\kf=20k_1$
in Model~A and $\kf=16k_1$ in Model~B.
An important control parameter of our simulations is the magnetic Reynolds
number, $\Rm=\urms/\eta\kf$, which is varied between 2 and 68,
although we also present a result with a larger value of $\Rm$.
This last simulation may not have run long enough and will therefore
not be analyzed in detailed.

We perform DNS of the equations of compressible  MHD
for an isothermal gas with constant sound speed $c_{\rm s}$,
\begin{eqnarray}
\label{mhd1}
\Dt\u &=& -c_{\rm s}^2\grad\ln\rho + \frac{1}{\rho}\JJ\times\BB 
 + \bm{F}_{\rm visc} + \f, \\
\Dt\ln\rho &=& -\grad\cdot\u, \\
\delt\AA &= & \u\times\BB - \eta\mu_0\JJ-\grad\psi,
\end{eqnarray}
where $\bm{F}_{\rm visc}=(\mu/\rho)(\lap\u + \frac{1}{3}\grad\dive\u)$
is the viscous force when the dynamic viscosity $\mu$ is constant (Model~A),
and $\bm{F}_{\rm visc}=\nu(\lap\u + \frac{1}{3}\grad\dive\u+2\SSSS\ln\rho)$
is the viscous force when the kinematic viscosity $\nu$ is constant (Model~B),
$\u$ is the velocity, 
$\JJ = \curl \BB/\mu_0$ is the current density,
$\mu_0$ is the vacuum permeability
(in the following we measure the magnetic field in Alfv\'en units
by setting $\mu_0=1$ everywhere),
$\rho$ is the density,
$\psi$ is the electrostatic potential,
and $D_t \equiv \delt + \u\cdot\grad$ is the advective derivative.
Here $\fxt$ is an external random white-in-time helical function of space 
and time.  
The simulations were performed with the {\sc Pencil Code}%
\footnote{{http://www.nordita.org/software/pencil-code/}},
which uses sixth-order explicit finite differences in space and third
order accurate time stepping method.
We use a numerical resolution of $128\times128\times256$ meshpoints.

\begin{figure*}[t!]
\begin{center}
  \includegraphics[width=\textwidth]{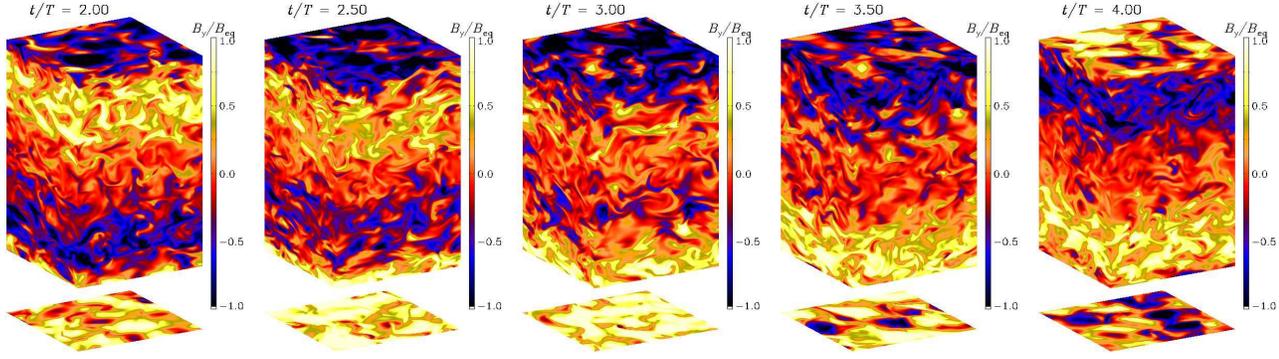}
\end{center}
\caption{Visualization of the $B_y$ component of the magnetic field on
the periphery of the domain at different times showing the migration of
magnetic patterns from the top and bottom boundaries toward the equator.
Time is measured in turbulent diffusion times, $T=(\etatz k_1^2)^{-1}$,
where $\etatz=\urms/3\kf$ is used as reference
}

\label{fig:bfly}
\end{figure*}

These simulations in a Cartesian box capture the essential aspects of
the simulations of \cite{mit+tav+kap+bra09} in spherical
wedge-shaped domains. In particular, in this case we also observe  
the generation of large-scale magnetic fields
which show oscillations on dynamical time scales, 
reversals of polarity and equatorward migration,
as can be seen from the sequence of snapshots in Fig.~\ref{fig:bfly}
for a run with $\Rm=68$.
Here we express time in units of the expected turbulent diffusion time,
$T=(\etatz k_1^2)^{-1}$, where $\etatz=\urms/3\kf$ is used as reference
value \cite{sur+bra+sub08}.

Below we shall employ this setup to study
the magnetic helicity and its flux. 
We shall discuss the issue of gauge-dependence in Sect.~\ref{gauge-dependence}.
\section{Magnetic helicity fluxes}
\label{formalism}
Let us first summarize the role played by magnetic helicity and its fluxes
in large-scale helical dynamos.  
The simplest case is that of a closed 
domain, i.e., one with periodic or perfect-conductor boundary conditions.
In the spirit of mean-field theory, we define large-scale
(or mean) quantities, denoted by an overbar, as a horizontal
average taken over the $x$ and $y$ directions. In addition, we denote a volume
average by angular brackets, $\brac{\cdot}$.
The magnetic helicity density is denoted by 
\begin{equation}
\hM \equiv \AA \cdot \BB,
\end{equation}
and its integral over a volume $V$ is denoted  by 
\begin{equation}
\HM \equiv \langle \hM \rangle \equiv \frac{1}{V}\int_V \AA \cdot \BB \,\dd V.
\end{equation}
In general the evolution equation of $\hM$ can be written down using
the MHD equations, which yields
\begin{equation}
\partial_t \hM  = -2 \EE \cdot \BB  - \dive \FFh,
\label{dhdt}
\end{equation}
where 
\begin{equation}
\FFh = \EE \times \AA + \psi \BB
\end{equation}
is the magnetic helicity flux
and $\EE$ is the electric field, which is given by 
\begin{equation}
\EE = -\u \times \BB + \eta \JJ.
\end{equation}
Given that our system is statistically homogeneous in the horizontal
directions, we consider the evolution equation for the horizontally
averaged magnetic helicity density,
\begin{equation}
\partial_t \hMM  = -2\eta\overline{\JJ \cdot \BB}  - \dive \FFhh,
\label{dhdtaver}
\end{equation}
where the contribution from the full electromotive force, $\u\times\BB$,
has dropped out after taking the dot product with $\BB$.
However, the mean electromotive force from the fluctuating fields,
$\meanEMF=\overline{\uu\times\bb}$, enters the evolution of the mean
fields, so this contribution does not vanish if we consider separately
the contributions to $\hMM$ that result from mean and fluctuating fields, i.e.
\begin{equation}
\partial_t \hMm = 2\meanEMF\cdot{\overline \BB} 
                  - 2\eta {\overline \JJ} \cdot {\overline \BB}
                    -\dive\FFhm,
\end{equation}
\begin{equation}
\partial_t \hMf = -2\meanEMF\cdot{\overline \BB} 
                  - 2\eta \overline{\jj\cdot\bb}
                    -\dive\FFhf,
\label{hMf}
\end{equation}
where 
\begin{eqnarray}
\FFhm &=& {\overline \EE} \times {\overline\AA}+\overline{\Psi}\,\overline{\BB},
\label{Fhm} \\
\FFhf &=&  \overline{\ee \times \aa} + \overline{\psi \bb},
\end{eqnarray}
and $\Psi=\overline{\Psi}+\psi$.

In mean-field dynamo theory one solves the evolution equation for
$\overline{\BB}$, so $\FFhm$ is known explicitly from the actual
mean fields.
However, the evolution equation for $\hMf$ is not automatically
obeyed in the usual mean-field treatment.
This is the reason why in the dynamical quenching formalism this
equation is added as an additional constraint equation.
The terms $\hMf$ and $\overline{\jj\cdot\bb}\approx\kf^2\hMf$ are
coupled to the mean-field equations through an additional contribution
to the $\alpha$ effect with a term proportional to $\kf^2\hMf$.
However, the coupling of the flux term $\FFhf$ is less clear, because
there are several possibilities and their relative importance is
not well established.

In this paper we are primarily interested in $\FFhf$ across the equator. 
We assume that this flux can be written in terms of the gradient
of the magnetic helicity density via a Fickian diffusion law, i.e.,
\begin{equation}
\FFhf = - \kappaf \grad \hMf,
\label{dflux}
\end{equation}
where $\kappaf$ is an effective diffusion coefficient for the magnetic
helicity density.

There are several points to note regarding Eq.~(\ref{dflux}). 
Firstly, both the magnetic helicity and its flux are {\it gauge-dependent}. 
Hence this expression should in principle depend on the gauge we choose.
However, as catastrophic quenching is a physically observable
phenomenon, it should not depend on the particular gauge chosen.
Secondly, we recall that Eq.~(\ref{dflux}) is purely a conjecture
at this stage, and it is the aim of this paper to test this conjecture.
Thirdly, Eq.~(\ref{dflux}) is not the only form of 
flux of magnetic helicity possible. Two other obvious 
candidates are the advective flux and the Vishniac--Cho
flux (\cite{vis+cho01}).
However, none of them can be of importance to the problem at hand, because
we have neither a large-scale velocity (thus ruling out advective flux) nor
a large-scale shear (thus ruling out Vishniac--Cho flux). 

\section{Diffusive flux and $\Rm$ dependence}
\label{results}

Let us postpone the discussion of the complications arising from the choice
of gauge until Sect.~\ref{gauge-dependence} and use the {\em resistive gauge}
for the results reported in this section, i.e.\ we set
\begin{equation}
\label{resistive}
\psi = \eta \dive\AA.
\end{equation}
We then calculate $\Fhf$ and $\hMf$ as functions of $z$ from our
simulations, time-average both of them and use Eq.~(\ref{dflux})
to calculate $\kappaf$ from a least-square fit of $\Fhf$ versus $-\nab\hMf$
within the range $-1.3\leq k_1z\leq 1.3$.
The values of $\kappaf$ as a function of $\Rm$ is
given in the last column of Table~\ref{TermsRHS}. 

\begin{figure}[h]
\begin{center}
  \includegraphics[width=\linewidth]{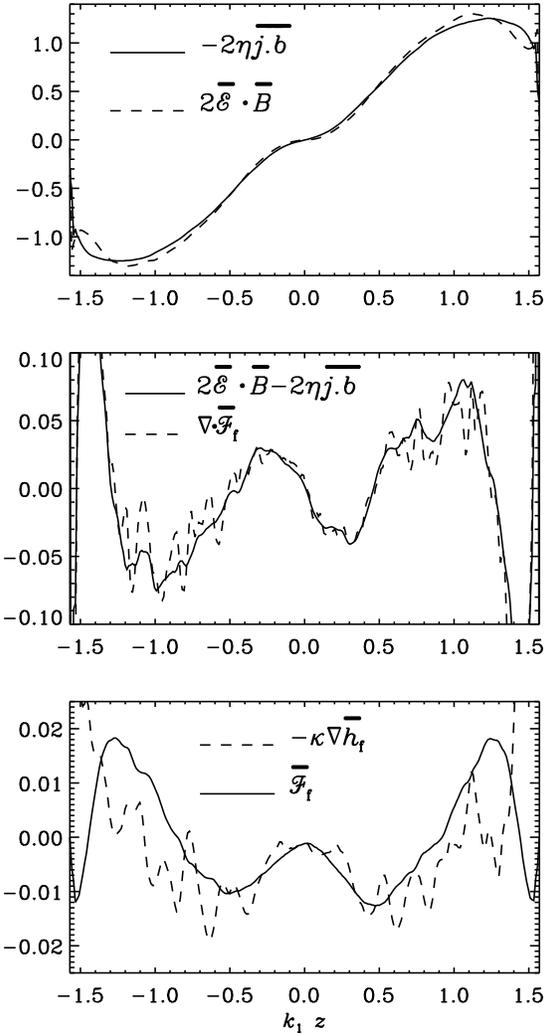}
\end{center}
\caption{
$z$ dependence of the terms on the RHS of Eq.~(\ref{hMf})
in the first two panels and in Eq.~(\ref{dflux}) for Run~B5.
}
\label{fig:pflux}
\end{figure}

In order to determine the relative importance of equatorial magnetic
helicity fluxes, we now consider individually the three terms
on the RHS of Eq.~(\ref{hMf}).
Within the range $-1.3\leq k_1z\leq 1.3$, all three terms vary
roughly linearly with $z$.
We therefore determine the slope of this dependence.
In Table~\ref{TermsRHS} we compare these three terms at $k_1z=-1$,
evaluated in units of $\etatz k_1\Beq^2$, as well as the
value of $\kappaf/\etatz$.
In Fig.~\ref{fig:pflux} we show the $z$ dependence of these three
terms for Run~B5, where $\Rm=68$.
The values of $\kappaf$ as a function of $\Rm$ is
given in the last column of Table~\ref{TermsRHS}. 
The $z$ dependence of $\Fhf$ and $\hMf$ is shown in the last panel of
Fig.~\ref{fig:pflux}.
Note that the two profiles agree quite well.

\begin{table}[t!]\caption{
Dependence of ${\overline\BB}^2$, normalized by $\Beq^2$,
the slopes of the three terms on the RHS of Eq.~(\ref{hMf}),
normalized by $\etatz\Beq^2$,
as well as the value of $\kappaf/\etatz$.
}\vspace{12pt}\centerline{\begin{tabular}{lccccccc}
Run & $\Rm$ & ${\overline\BB}^2$
& $2\meanEMF\cdot\overline \BB$ & $2\eta\overline{\jj\cdot\bb}$
& $\dive\FFhf$ & $\kappaf/\etatz$\\
\hline
B1 &  2 & 1.1 &  9.42 & $-9.38$ & $-0.04$ & 0.41 \\
B2 &  5 & 2.2 & 11.18 &$-11.14$ & $-0.04$ & 0.34 \\
B3 & 15 & 2.0 &  4.54 & $-4.52$ & $-0.02$ & 0.27 \\
B4 & 33 & 1.7 &  2.28 & $-2.27$ & $-0.01$ & 0.31 \\
B5 & 68 & 0.8 &  1.15 & $-1.12$ & $-0.03$ & 0.34 \\
\label{TermsRHS}\end{tabular}}\end{table}

We point out that, near $z=0$, all simulations show either a local
reduction in the gradients of the terms on the RHS of Eq.~(\ref{hMf})
or even a local reversal of the gradient.
This is likely to be associated with a local reduction in dynamo activity
near $z=0$, where kinetic helicity is zero.
The non-uniformity of the turbulent magnetic field also leads to
transport effects \cite{BS05} that may modify the gradient.
However, we shall not pursue this question further here.

\begin{figure}[h]
\begin{center}
  \includegraphics[width=\linewidth]{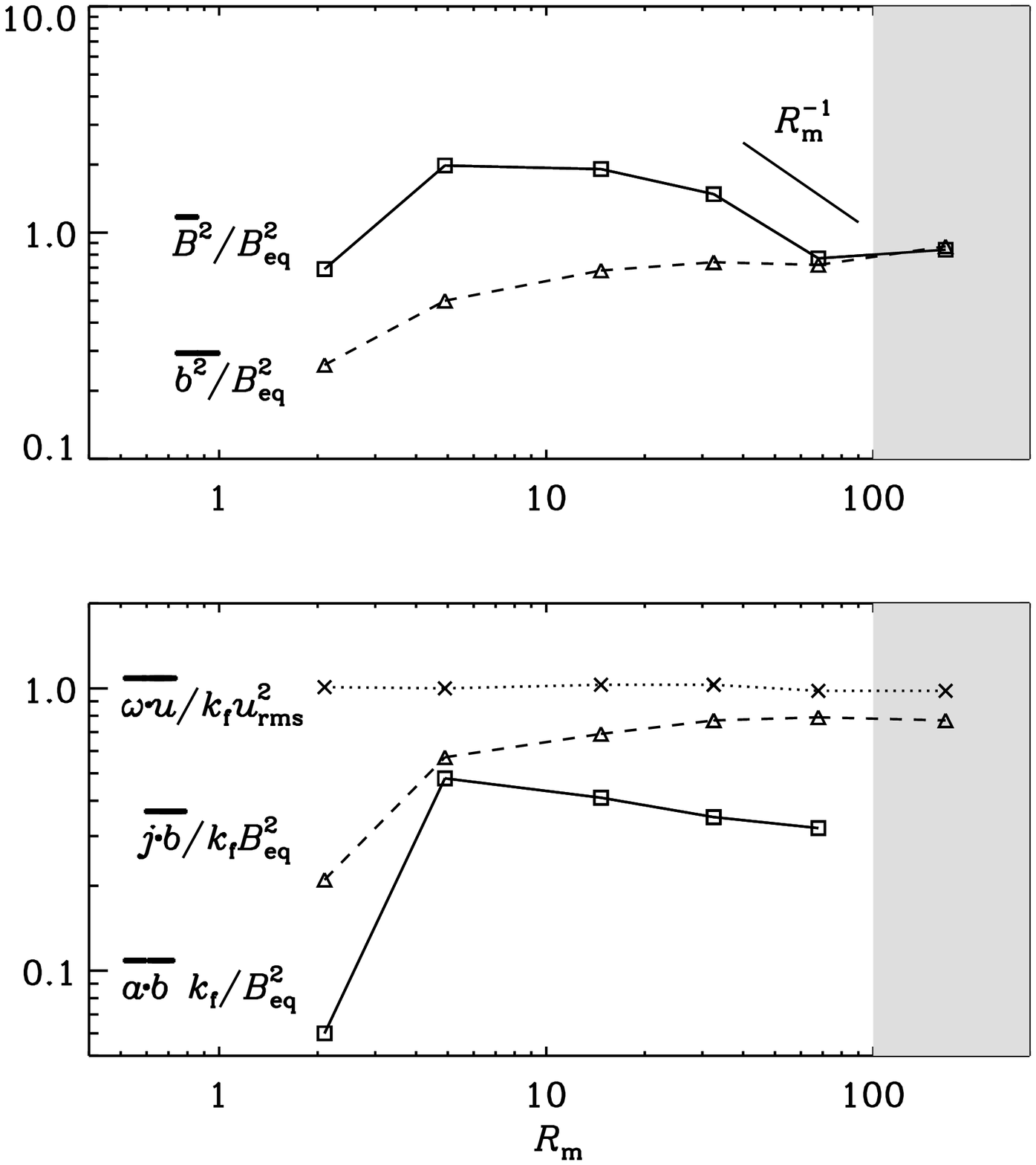}
\end{center}
\caption{
$\Rm$ dependence of the normalized magnetic energy of the mean field,
$\bra{\overline\BB^2}/\Beq^2$,
and the fluctuating field, $\bra{\bb^2}/\Beq^2$, in the upper panel
together with the normalized helicities of
the small-scale magnetic field, $\overline{\aa\cdot\bb}\kf/\Beq^2$,
the small-scale current density, $\overline{\jj\cdot\bb}/\kf\Beq^2$,
and the small-scale velocity, $\overline{\oo\cdot\uu}/\kf\urms^2$,
at $k_1z=-1$ (i.e.\ in the south) in the lower panel.
(All three helicities are negative in the north and positive in the south.)
The shaded ares indicates that the solutions are different in nature,
and that the simulations may not have run for long enough.
}
\label{fig:pbsat2}
\end{figure}

Looking at Table~\ref{TermsRHS}, we see that $2\meanEMF\cdot\overline \BB$
and $2\eta\overline{\jj\cdot\bb}$ balance each other nearly perfectly,
and that only a small residual is then balanced by the diffusive flux
divergence, $\dive\FFhf$.
For the values of $\Rm$ considered here, the terms
$2\meanEMF\cdot\overline \BB$ and $2\eta\overline{\jj\cdot\bb}$
scale with $\Rm$, while the dependence of $\dive\FFhf$ on $\Rm$
is comparatively weak.
If catastrophic quenching is to be alleviated by the magnetic helicity
flux, one would expect that at large values of $\Rm$ the terms
$2\meanEMF\cdot\overline \BB$ and $\dive\FFhf$ should balance.
At the moment our values of $\Rm$ are still too small by about
a factor of 30-60 (assuming that the same scaling with  $\Rm$ persists).
This result is compatible with that of earlier mean field models (\cite{BCC09}).
Consequently, we see that the energy of the mean magnetic field
decreases with increasing from $\Rm=33$ to 68; see Fig.~\ref{fig:pbsat2}.
For larger values of $\Rm$ the situation is still unclear.

In Table~\ref{TermsRHS}, we also give the approximate values of
$\kappaf/\etatz$.
Note that this ratio is always around 0.3 and independent of $\Rm$.
This is the first time that an estimate for the diffusion coefficient
of the diffusive flux has been obtained.
There exists no theoretical prediction for value of $\kappaf$ other than
the naive expectation that such a term should be expected and that its
value should be of the order of $\etatz$.
This now allows us to state more precisely the point where the turbulent
diffusive helicity flux becomes comparable with the resistive term,
i.e.\ we assume $\kappaf\nabla^2\overline{\aa\cdot\bb}$ to become
comparable with $2\eta\overline{\jj\cdot\bb}$.
Using the relation $\overline{\jj\cdot\bb}\approx\kf^2\overline{\aa\cdot\bb}$
\cite{BB02}, which is confirmed by the current simulations within a factor
of about 2 (see the second panel of Fig.~\ref{fig:pbsat2}), we find that
\begin{equation}
\kappaf/2\eta > (\kf/k_1)^2,
\end{equation}
where we have assumed that the Laplacian of $\overline{\aa\cdot\bb}$
can be replaced by a $k_1^2$ factor.
Using our empirical finding, $\kappaf\approx\etatz/3$, together with
the definition $\etatz/\eta\approx\urms/3\eta\kf=\Rm/3$, we arrive at
the condition
\begin{equation}
\Rm>18(\kf/k_1)^2\approx4600 \quad\mbox{(for $\kappaf$ to be important)},
\end{equation}
where we have inserted the value $\kf/k_1=16$ for the present simulations.
Similarly, large values of $\Rm$ for alleviating catastrophic quenching by
turbulent diffusive helicity fluxes were also found using mean-field
modelling (\cite{BCC09}).
Unfortunately, the computing resources are still not sufficient to
verify this in the immediate future.

\section{Gauge-dependence of helicity flux}
\label{gauge-dependence}

\begin{figure}[h]
\begin{center}
  \includegraphics[width=.80\linewidth]{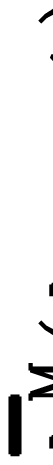}
\end{center}
\caption{ Plot of $\hMf$ as a function of time in the statistically
stationary state for $k_1 z = -1$ (south, top panel) and 
$k_1 z = 1$ (north, bottom panel) for the three different gauges,
Weyl gauge (open circle), Lorenz gauge (line) and resistive gauge (broken
line).  }
\label{fig:ts}
\end{figure}

Let us now consider the question of gauge-dependence of the
helicity flux. Equation~(\ref{hMf})
is obviously gauge-dependent.
However, if, in the statistically steady state, $\hMf$ becomes
independent of time, we can average this equation and obtain
\begin{equation}
{\partial\Fhf\over\partial z}=-2\meanEMF\cdot\overline \BB
-2\eta\overline{\jj\cdot\bb},
\label{hMf0}
\end{equation}
where $\Fhf$ refers to the $z$ component of $\FFhf$.
On the RHS of this equation the two terms are gauge-independent.
Therefore $\dive \Fhf$ must also be gauge-independent.
The same applies also to $\Fhm$ and $\overline \Fh$; see Eq.~(\ref{dhdtaver}).
We have confirmed that, in the steady state, $\hMf$ is statistically
steady and does not show a long-term trend;
cf.\ Fig.~\ref{fig:ts} for the three gauges. We note that the fluctuations
of $\hMf$ are typically much larger for the Weyl gauge than for the other two.  

We now verify the expected gauge-independence explicitly for three different 
gauges: 
\noindent the {\em Weyl gauge}, defined by 
\begin{equation}
\psi=0,
\end{equation}
the {\em Lorenz gauge} (or pseudo-Lorenz gauge)\footnote{In fact, this
is not the true Lorenz gauge
because we use velocity of sound \cite{BK07} instead of the velocity of
light which appears in the original Lorenz gauge}, defined by
\begin{equation}
\partial_t \psi = -c^2_{\psi}\dive\AA,
\label{restive-gauge}
\end{equation}
and the resistive gauge, defined by (\ref{resistive}) above.
We calculate the normalized magnetic helicity for both the 
mean and fluctuating parts and the respective fluxes 
for all the three gauges. These simulations are done
for the Model~A with low $\Rm$ ($\Rm \approx 1.9$).

\begin{figure}[h]
\begin{center}
  \includegraphics[width=.80\linewidth]{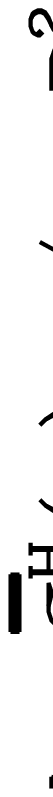}
\end{center}
\caption{
Comparison of the flux $\Fhf(z,t)$ at a randomly chosen instant (upper panel)
and its time average $\Fhf(z)$ for the three different gauges.
Lorenz gauge ($\circ$), Weyl gauge ($\diamond$) and the 
resistive gauge ($\cdot$). The instantaneous flux is plotted 
in the top panel and the time-averaged flux is plotted in the bottom 
panel. }
\label{fig:ffhf}
\end{figure}

We find the transport coefficient $\kappaf$ in the way 
described in the previous section. 
A snapshot of the mean flux $\Fhf$ is plotted in the top
panel of Fig.~\ref{fig:ffhf}. The flux is different in all
the three gauges. 
However, when averaged over the horizontal directions as well as
time the fluxes in the three different
gauges agree with one another as shown in the bottom panel of
Fig.~\ref{fig:ffhf}. 
We find the transport coefficient $\kappaf$ as described in the
previous section and obtain the same value in all the three gauges.

\section{Conclusion}
\label{conclusion}
In this paper we use a setup in which the two parts
of the domain have different signs of kinetic and
magnetic helicities. Using DNS we show that the flux of magnetic helicity
due to small-scale fields can be described by Fickian diffusion down
the gradient of this quantity.
The corresponding diffusion coefficient is approximately independent of $\Rm$.
However, in the range of $\Rm$ values we have considered here,
the flux is not big enough to alleviate the catastrophic quenching.
The critical value of $\Rm$ for the flux to become important is proportional
to the square of the scale separation ratio.
In the present case, where this ratio is 16, the critical value of $\Rm$
is estimated to be 4600.
We have also calculated the flux and the diffusion 
coefficient in the three gauges discussed above and have
found the fluxes to be independent of the choice of these gauges.
This is explained by the fact that in the steady state the 
divergence of magnetic helicity flux 
is balanced by terms that are gauge-independent.

Several immediate improvements on this study spring to mind.
One is to compare our results with the gauge-independent magnetic helicity
of Berger \& Field (1984) and the corresponding magnetic helicity flux.
The second is to extend the present study to
higher values of $\Rm$ to understand the asymptotic behavior
of the flux.  
Finally, it may be useful to compare the results for different
profiles of kinetic helicity to see whether or not our results
depend on such details.

\acknowledgements{The simulations were performed with the computers 
hosted by QMUL HMC facilities purchased under the SRIF initiative.
We also acknowledge the allocation of computing resources provided by the
Swedish National Allocations Committee at the Center for
Parallel Computers at the Royal Institute of Technology in
Stockholm and the National Supercomputer Centers in Link\"oping.
This work was supported in part by
the European Research Council under the AstroDyn Research Project 227952
and the Swedish Research Council grant 621-2007-4064.
DM is supported by the Leverhulme Trust.}




\begin{thebibliography}{9}
\expandafter\ifx\csname natexlab\endcsname\relax\def\natexlab#1{#1}\fi

\bibitem[{(Berger 1984)}]{Berger84}
Berger, M.: 1984, GApFD 30, 79

\bibitem[{Berger \& Field 1984}]{ber+fie84}
Berger, M., Field, G.B.: 1984, JFM 147, 133

\bibitem[{Blackman \& Field 2000}]{BF00}
Blackman, E.G., Field, G.B.: 2000, \apj 534, 984

\bibitem[{(Blackman \& Brandenburg 2002)}]{BB02}
Blackman, E.G., Brandenburg, A.: 2002, \apj 579, 359

\bibitem[{(Brandenburg \& K\"apyl\"a 2007)}]{BK07}
Brandenburg, A., K\"apyl\"a, P.J.: 2007, New J. Phys. 9, 305

\bibitem[{(Brandenburg \& Subramanian 2005a)}]{BS05}
Brandenburg, A., Subramanian, K.: 2005, PhR 417, 1

\bibitem[{({Brandenburg} \& {Subramanian} 2005b)}]{BS05c}
{Brandenburg}, A. \& {Subramanian}, K.: 2005, AN 326, 400

\bibitem[{{Brandenburg} et al.\ 2009}]{BCC09}
{Brandenburg}, A., {Candelaresi}, S., {Chatterjee}, P.: 2009, \mnras 398, 1414

\bibitem[{(Brandenburg et al.\ 2008)}]{BRRK08}
Brandenburg, A., R\"adler, K.-H., Rheinhardt, M., K\"apyl\"a, P.J.: 2008,
\apj 676, 740

\bibitem[{{Covas} et al.\ 1998}]{CTTB98}
Covas, E., Tavakol, R., Tworkowski, A., Brandenburg, A.: 1998, A\&A 329, 350

\bibitem[{{Kleeorin} \& {Rogachevskii} 1999}]{KR99}
Kleeorin, N., Rogachevskii, I.: 1999, Phys Rev E 59, 6724

\bibitem[{{Kleeorin} et al.\ 2000}]{KMRS00}
Kleeorin, N., Moss, D., Rogachevskii, I., Sokoloff, D.: 2000, A\&A 361, L5

\bibitem[{{Mitra} et al.\ (2009)}]{mit+tav+kap+bra09}
{Mitra}, D., {Tavakol}, R., {K{\"a}pyl{\"a}}, P.J., {Brandenburg}, A.: 2009,
  arXiv:0901.2364

\bibitem[{Shukurov et al.\ 2006}]{SSSB06}
Shukurov, A., Sokoloff, D., Subramanian, K., Brandenburg, A.: 2006,
A\&A 448, L33

\bibitem[{{Subramanian} \& {Brandenburg} 2004}]{SB04}
{Subramanian}, K., {Brandenburg}, A.: 2004, Phys Rev Lett 93, 205001

\bibitem[{{Subramanian} \& {Brandenburg} 2006}]{SB06}
{Subramanian}, K. {Brandenburg}, A.: 2006, \apjl 648, L71

\bibitem[{({Sur} et al.\ 2008)}]{sur+bra+sub08}
{Sur}, S., {Brandenburg}, A., {Subramanian}, K.: 2008, \mnras 385, L15

\bibitem[{{Vishniac} \& {Cho} 2001}]{vis+cho01}
{Vishniac}, E.T., {Cho}, J.: 2001, \apj 550, 752

\end{thebibliography}
\end{document}